\documentclass[a4paper,11pt]{article}
\pdfoutput=1 

\usepackage{jheppub} 

\usepackage[T1]{fontenc} 
\usepackage{ulem}
\usepackage{color}
\usepackage{graphicx}
\usepackage{hyperref}   
\usepackage{appendix}
\usepackage{epsfig}
\usepackage{epstopdf}
\usepackage{amsmath,amssymb}
\usepackage{slashed}
\usepackage{csquotes}
\usepackage{caption}
\usepackage{subcaption}
\usepackage{comment}
\usepackage[dvipsnames]{xcolor}

\title{\boldmath Non-perturbative diffusion of Heavy Quark moving in a hot and magnetised Quark Gluon Plasma}

\author[a,1]{Surasree Mazumder,\note{Corresponding author.}}
\author[a]{Vinod Chandra,}
\author[b]{Santosh K Das}

\affiliation[a]{Indian Institute of Technology, Gandhinagar,\\Gujarat-382055, India}
\affiliation[b]{Indian Institute of Technology, Goa\\Ponda-403401, Goa, India}

\emailAdd{surasree.m@iitgn.ac.in}
\emailAdd{vchandra@iitgn.ac.in}
\emailAdd{santosh@iitgoa.ac.in}

\abstract{Heavy Quarks (HQs) serve as excellent probes to understand various characteristics of deconfined hot QCD medium, comprising light quarks and gluons, created in the Heavy Ion Collisions (HICs).  Strong magnetic fields in non-central HICs may significantly affect HQ dynamics in this medium.
Exploring the impact of the magnetic field on the perturbative and non-perturbative transport coefficients of HQ is an intriguing endeavor. This necessitates the development of a comprehensive theoretical framework accommodating the non-perturbative Quantum Chromo Dynamics(npQCD) description alongside perturbative QCD(pQCD). The present work provides such a formulation in which Quarkonium potential in the hot and magnetic QCD medium is implemented as the effective gluon propagator to calculate the rate of elastic scattering between HQ and the light partons inside a medium of hot Quark Gluon Plasma(QGP) in the presence of a strong but uniform magnetic field. Diffusion coefficients of a charm quark have been computed for a short-range Yukawa potential ( pQCD) as well as a long-range confining/non-perturbative potential (npQCD) for two cases in which the velocity of the charm quark is parallel and perpendicular to the magnetic field respectively. Non-perturbative contribution is seen to dominate over the perturbative one in the regime of low temperatures and low to intermediate momenta of charm quark. As the momentum of charm quark as well as the temperature of the medium increase pQCD starts to gain on the npQCD contribution, ultimately prevailing  at higher temperatures and charm momenta.
}

\begin{document} 
\maketitle
\flushbottom

\section{Introduction}
\label{introduction}

Heavy quarks (HQs)~\cite{Prino:2016cni,Andronic:2015wma,Rapp:2018qla,Aarts:2016hap,Cao:2018ews,Dong:2019unq,Das:2022lqh}  like charm and bottom are considered as excellent probes to study the properties of hot QCD matter (commonly known as Quark-Gluon-Plasma (QGP)) produced in relativistic heavy-ion collisions (RHIC). As they are produced at a very early stage of the collisions, well before the formation of the thermal medium (QGP), they witness the entire space-time evolution of the  QGP medium until the freeze-out and thus can help understand the bulk and transport properties of the QGP through HQ observable  concerning collective flow and energy loss. Due to its heavy constituent mass, HQ motion can be described by the equations of a Brownian particle, namely Langevin or Fokker Planck equation, immersed in a fluid comprising light particles.
A plethora of studies formulated within the framework of Brownian motion has estimated various transport coefficients of HQ interacting with the light quarks and gluons of the hot medium. 

Recent experimental observations reveal the creation of a very large net magnetic field in non-central Heavy Ion Collisions due to the presence of spectator protons. The strength of the magnetic field, thus produced, can mount up to $eB\sim m_{\pi}^2$ in RHIC(relativistic Heavy Ion Collider) at BNL and $eB\sim 10-20~m_{\pi}^2$ in LHC(Large Hadron Collider) a CERN~\cite{IJMP24:2009,PRC83:2011_1,PLB710:2012,PRC85:2012,PLB718:2013,PRL110:2013}. Although the field starts very large at the onset of the collision, it decays fast. Therefore, whether such a field would survive till the thermalized medium (QGP) is formed depends on numerous properties of the early-stage or the pre-equilibrium medium. With a requisite amount of electrical conductivity of the medium, the decreasing magnetic field would induce a current that would, in congruence with the Lenz's law, extend the lifetime of the magnetic field and drag it into the thermal medium~\cite{PRC88:2013,NPA929:2014,PRC93:2016}. Under this circumstance, several interesting effects of the magnetic field on the properties of QGP have been explored, such as Chiral Magnetic Effect(CME)~\cite{NPA803:2008,PRD78:2008}, charge-dependent elliptic flow~\cite{PRL107:2011,PRD83:2011,PRC83:2011_2}, anisotropic production of photons and dileptons~\cite{PLB710-230:2012,PRL109:2012,PRD86:2012,PRD88:2013_1,PRD89:2014_1,PRC90:2014_1} and so forth. 

The strong magnetic field produced in heavy-ion collisions exerts a significant influence on the dynamics of the heavy quarks in QGP, as demonstrated in various studies~\cite{Das:2016cwd,STAR:2019clv,ALICE:2019sgg}. These studies may play crucial role in developing a deeper understanding of the intricate interplay between heavy quarks and magnetic fields in the context of the hot QCD medium. 
Recent research works have successfully ventured into the  estimation of the HQ transport coefficients in presence of a magnetic field in the hot QCD medium; ranging from the static properties of HQ~\cite{PRD88:2013_2,PRD89:2014_2,PRD93:2016_1} to the real-time dynamic properties~\cite{PRD93:2016_2,PRD100:2019,Kurian:2020kct,JHEP05:2020,arxiv2004:2020,PRD105:2022}.  Although the perturbative Quantum Chromo Dynamics(pQCD) has been successful in explaining the experimental observables such as the nuclear modification factor, $R_{AA}$ and elliptic flow, $v_2$, at higher $p_T$ of HQ, pQCD has not quite succeeded to elucidate the experimental data at low to intermediate $p_T$ region~\cite{Rapp:2018qla,JHEP02:2008}.  Numerous theoretical studies have been conducted to shed light on this complex domain, devising ways to calculate the HQ transport coefficients in the non-perturbative regime~\cite{vanHees:2005wb,PRL100:2008,Gossiaux:2008jv,Gossiaux:2009mk,Das:2012ck,PRC86:2012,Berrehrah:2013mua,PLB747:2015,Scardina:2017ipo,Song:2015sfa,Beraudo:2014boa,EPJC78:2018_1,EPJC78:2018_2,PRC101:2020,PRC94:2016,Xu:2017obm}. 

The T-matrix approach to calculate the HQ transport coefficients at low temperatures(closer to the critical temperature, $T_c$) has refined the method by making use of the interaction potential elicited from the finite-temperature lattice QCD studies and thus has taken a successful step towards computing the heavy quark transport coefficients in the low $p_T$ region~\cite{PRL100:2008,PRC86:2012, Liu:2018syc}.  A recent study~\cite{gribov:najmul} estimated the diffusion coefficient of a non-relativistic HQ using the Gribov-Zwanzinger action. Another approach has been proposed in Ref.\cite{arxiv2112:2021} in which a parameterized Cornell type HQ potential is assumed to work as an effective gluon propagator for the interaction of HQ with the medium partons to calculate HQ transport coefficients. The short-range Yukawa term is attributed to the pQCD description, and the long-range string or confinement term gives rise to the non-perturbative counterpart of the calculation. 

It is evident from the above discussion that the response of the medium due to the existence of the magnetic field and the non-perturbative nature of the QCD medium are two very crucial aspects that can affect HQ transport in diverse ways. Therefore, it is constructive to put together a formalism that can congregate these two salient features pertaining to HQ transport coefficients evaluated in presence of a magnetic field in the non-perturbative regime. In the present paper, we have attempted to develop this nascent idea into the real-time calculation of the diffusion coefficients of a charm quark moving with a velocity inside QGP with a magnetic field in the background. HQ potential inside a hot and magnetic plasma is complex in nature and has been calculated in Ref.\cite{PRD97:2018} in presence of a strong magnetic field obeying the specified scale hierarchy $M^2\gg eB\gg T^2$. Using this form of HQ potential, mapped into the momentum space, we calculate various diffusion coefficients of HQ in the pQCD and non-pQCD domains and discuss the associated implications.  

The paper is organized into four sections as follows.  Sec.\ref{formalism} is composed of four different subsections discussing, respectively, how the HQ potential in a hot QGP in presence of a strong magnetic field would resemble in the momentum space, the method to calculate the scattering rates of a relativistic HQ elastically interacting with the medium particles and calculations of the diffusion of HQ in the scenarios when the velocity of HQ is along the direction of the magnetic field as well as perpendicular to the field. Results and discussions regarding the diffusion coefficients are presented in Sec.\ref{Results}. We conclude in Sec.\ref{Summary} in which we present a gist of the work and share some future outlooks.  

\section{Formalism}
\label{formalism}

The physics of npQCD is instrumental in explaining the phenomenology of HQ at Relativistic HICs, specifically, at low temperatures and low momentum of HQ.
npQCD requires the QCD coupling to assume a far larger value than used in pQCD calculations, rendering it difficult to compute any physical quantity. Therefore, one has to avail oneself of some theoretical model encompassing all the intricacies of npQCD.
A possible way to address the problem from a particular standpoint where HQ-anti quark potential is used as the effective gluon propagator for all the calculations of the scattering processes of HQ with the medium partons. It is beneficial to emphasize the fact that whereas the Yukawa part of the potential is ascribed to the pQCD processes, the string part is responsible for the non-perturbative contribution.  

We start by writing down the HQ potential worked out in Ref.~\cite{PRD97:2018} mapped into momentum space inside a hot and strongly magnetized medium. Then we calculate the scattering rate of HQ interacting with the light quark/anti-quark and the gluons of the QGP medium and subsequently estimate HQ diffusion coefficients in the scenarios in which i) HQ moves along the direction of the magnetic field (along the z-axis) and ii) HQ velocity is perpendicular to the magnetic field. Axiomatically, all these calculations are performed in the limit $M^2\gg eB\gg T^2$.
\subsection{Heavy quark potential in hot QGP in presence of a strong magnetic field in momentum space}
The generic form of the Cornell type parametrized potential for Heavy Quark (HQ)-QGP scattering in a hot QCD medium without the presence of the magnetic field is
\begin{equation}
V(r)=V_Y(r)+V_S(r)=-\frac{4}{3}\alpha_s\frac{e^{-m_Dr}}{r}-\frac{\sigma (e^{-m_S r}-1)}{m_S}.
\label{Vrnomag}
\end{equation}
The expression in Eq.\ref{Vrnomag} comprises two parts, one is the short-range Yukawa potential designated by $V_Y$ and another is the long range color confinement potential or the string potential, $V_S$. It can be stipulated that this potential can act as an effective gluon propagator for all intents and purposes in the forthcoming calculations. 

The complex HQ potential inside a hot QGP has been calculated in Ref.~\cite{PRD97:2018} in presence of a strong magnetic field assuming the light quarks only occupy the Lowest Landau Levels(LLLs) with the approximation $eB\gg T^2$.
In order to use this potential as a gluon propagator, one requires it to be written down in momentum space. 
The differential equation followed by the Yukawa potential in coordinate space:
\begin{equation}
-\nabla^2 V_Y(r)+m_D^2 V_Y(r)=4\pi\alpha_s\delta(r)-i\alpha_s T m^2_{Dg}g(m_D r)
\label{difeqVcr}
\end{equation}
in which $m_{Dg}$ is the screening mass due to the thermal gluons present in the medium given by $m^2_{Dg}=\frac{4}{3}\alpha_sT^2N_c$ and $g(y)$ is an integral given by $g(y)=2\int_{0}^{\infty}\frac{x}{x^2+1}\frac{\sin{xy}}{xy}dx$.
Performing Fourier transform on both sides of Eq.\ref{difeqVcr}, we get
\begin{equation}
q^2V_Y(q)+m_D^2V_Y(q)=4\pi\alpha_s-4i\pi^2\alpha_s T m^2_{Dg}\frac{1}{q(q^2+m_D^2)}\nonumber
\end{equation}
which in turn can be written as 
\begin{equation}
V_Y(q)=4\pi\alpha_s\left[\frac{1}{(q^2+m_D^2)}-i\pi T m^2_{Dg}\frac{1}{q(q^2+m_D^2)^2}\right]
\label{VcqB}
\end{equation}
in which $m_D$ is the Debye mass in presence of a strong magnetic field, given by $m_D^2=m^2_{Dg}+\Sigma_{f}\frac{|q_f|B\alpha_s}{2\pi}$, where $f$ is the quark flavor in the quark loop in gluon self-energy.

It is illuminating to look at this problem from another perspective. It can be shown that the same expression(Eq.~\ref{VcqB}) for the Yukawa potential is arrived at starting from the generalized Gauss's law. Medium modified generalized Gauss law can be written as
\begin{equation}
q^2V_Y(q)=\frac{4\pi\alpha_s}{\epsilon(\Vec{q},m_D)},
\label{gauss}
\end{equation}
in which $\epsilon$ is the in-medium complex permittivity in presence of a finite temperature $T$ and strong magnetic field, $\Vec{B}$ given by
\begin{equation}
\epsilon^{-1}(\Vec{q},m_D)=\frac{q^2}{q^2+m_D^2}-i\pi Tm^2_{Dg}\frac{q}{(q^2+m_D^2)^2}.
\label{permittivity}
\end{equation}
A combination of Eq.\ref{permittivity} and Eq.\ref{gauss} can also give us Eq.\ref{VcqB}.

However, it is not trivial to derive an expression for the string part of the potential following a similar strategy as done in the case of the Yukawa one. It is not feasible to use the Gauss law method to write down the string potential. One can begin with the differential equation describing the string potential, $V_S(r)$, and then Fourier transform it to obtain an equation in momentum space. The differential equation is given by
\begin{equation}
-\frac{1}{r^2}\frac{d^2V_S(r)}{dr^2}+m_s^4V_S(r)=4\pi\sigma\delta(r)-i\sigma T m_{Dg}^2g(m_Dr).
\label{diffeqVsr}
\end{equation}
Taking Fourier transform of Eq.\ref{diffeqVsr}, we get the following equation:
\begin{equation}
-4\pi \sigma-q^2\frac{d^2\Bar{V_S(q)}}{dq^2}+m_S^4V_S(q)=4\pi\sigma\left[1-\frac{i\pi T m_{Dg}^2}{q(q^2+m_D^2)}\right],
\label{ftdiffeqvsr}
\end{equation}
where, $\Bar{V_S(q)}=F.T\left[\frac{V_S(r)}{r^4}\right]$ and $m_S^2=(\frac{m_{Dg}^2\sigma}{\alpha_s})^{1/2}$.  Eq.\ref{ftdiffeqvsr} cannot be reduced to an equation involving only $V_S(q)$, and yet our objective is to find such an equation after all. 
So, we adopt a roundabout way by observing a few basic prerequisites that $V_S(q)$ must have.
From the form of the string potential in the absence of the magnetic field (Eq.\ref{Vrnomag}), it is easy to observe that the Fourier transform of the string part is proportional to $\sigma/(q^2+m_s^2)^2$. 
If one notices the similarity between the last terms of the right-hand sides of Eq.~\ref{difeqVcr} and Eq.~\ref{diffeqVsr}, one can be sure that the Fourier transform of $g(m_Dr)$ should appear in the final equations of both $V_Y(q)$ and $V_S(q)$ in the same manner. One can also notice that the last term of Eq.\ref{VcqB} depends on the square of the coupling and therefore the last term of $V_S(q)$ should depend on the square of the string tension. On account of these requirements, the form of the string potential in momentum space can be conjectured to be:
\begin{equation}
V_S(q)=4\pi\sigma\left[\frac{2}{(q^2+m_s^2)^2}-\frac{i\pi \sigma T m^2_{Dg}}{q(q^2+m_D^2)(q^2+m_s^2)^3}\right].
\label{VsqB}
\end{equation}

\subsection{Scattering rate of HQ in the medium of light partons}

Let us consider the elastic scattering of HQ with the light quarks in LLLs and the thermal gluons. By cutting the diagram representing HQ self-energy, one gets two elastic scattering processes: 1) Coulombic process which is the elastic scattering of HQ with the light quarks/anti-quarks and 2) Compton process which is the elastic scattering of HQ with the thermal gluons.

\begin{figure}[h]
\includegraphics[width=0.8\textwidth]{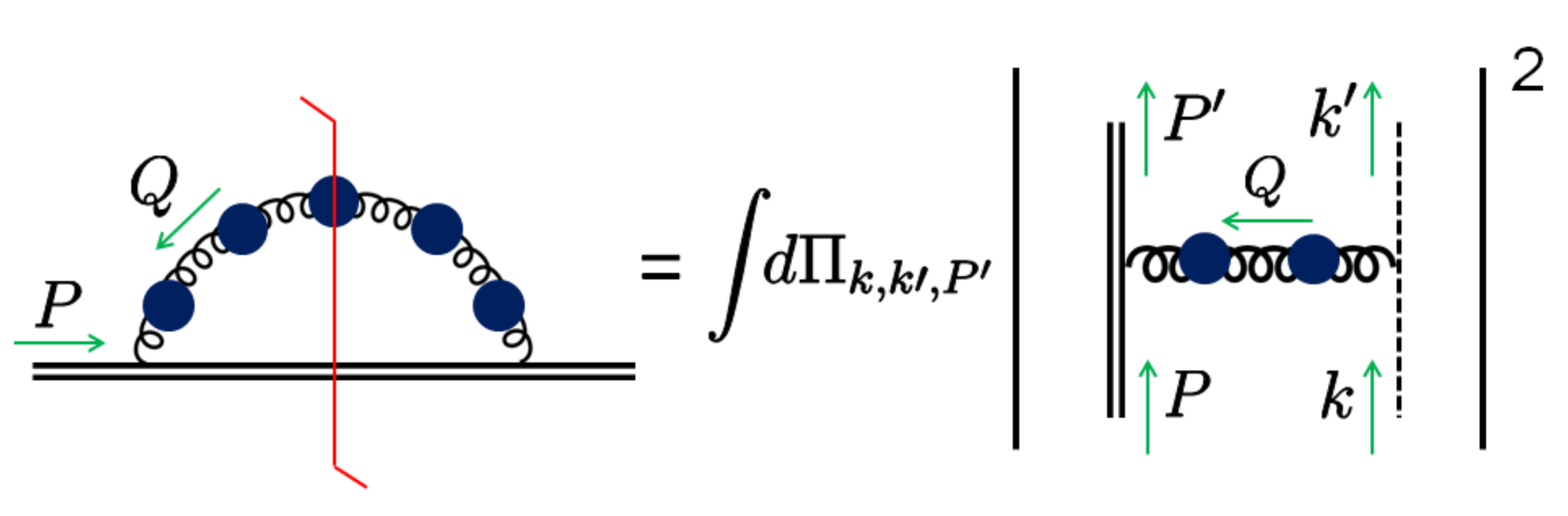}
\caption{HQ self-energy with the effective gluon propagator cut into two elastic scattering processes~\cite{PRD93:2016_2}}
\label{fig1}
\end{figure}

The scattering rate of HQ with an energy $E$ and momentum $\Vec{p}$ in the QCD medium is given by the following expression:
\begin{equation}
\Gamma(E)=-\frac{1}{2E\left(1+e^{-E/T}\right)}Tr\left[(\slashed{P}+M)\Im{\Sigma(P)}\right],
\label{scatrate}
\end{equation}
in which $\Sigma(P)$ is the self energy of HQ. The effective gluon propagator is the HQ potential as discussed in the previous section.

\subsubsection{Scattering rate due to Yukawa part of HQ potential}

One can calculate the self-energy of HQ from the following expression:
\begin{equation}
\Sigma(P)=ig^2\int \frac{d^4Q}{(2\pi)^4}\gamma_{\mu}\frac{1}{\slashed{P'}-M}\gamma^{\mu}V_Y(\Vec{q}),
\label{HQselfc}
\end{equation}
where $1/(\slashed{P'}-M)$ is the HQ propagator unaffected by the presence of the magnetic field due to its heavy mass.

To compute the four-dimensional integral using imaginary time formalism, one replaces $\int\frac{dq_0}{2\pi}\rightarrow iT\sum_{q_0}$, in which case the sum is over the discrete imaginary values of $q_0=2ni\pi T$ for the gluon energy. The summation over $q_0$ is performed for discrete imaginary values $p_0=(2n+1)i\pi T$ for the HQ energy. Only in this scenario, $p_0$ can analytically be continued to real Minkowski energy, $p_0=E+i\epsilon$. 
Implementing the above prescription, one can write
\begin{align}
Tr\left[(\slashed{P}+M)\Sigma(P)\right]&=-g^2T\sum_{q_0}\int \frac{d^3q}{(2\pi)^3}\frac{V_Y(q)}{P'^2-M^2}\nonumber\\
&\times Tr\left[(\slashed{P}+M)\gamma_{\mu}(\slashed{P'}+M)\gamma^{\mu}\right].
\label{traceC1}
\end{align}

Plugging in Eq.\ref{traceC4}, Eq.\ref{rhoY} and Eq.\ref{VYqalter} into Eq.\ref{scatrate} and ignoring terms $\sim$ $T^2/q^2$, one obtains the expression for the scattering rate of HQ with the light partons of the medium in case of a Yukawa type interaction as
\begin{align}
&\Gamma_Y(E)=\frac{2\pi g^2M^2}{E^2}\int\frac{d^3q}{(2\pi)^3}\rho_Y(q)\int d\omega\nonumber\\ 
& [1+n_B(\omega)]\delta(\omega-\Vec{v}\cdot\Vec{q}).
\label{scatrateC}
\end{align}

\subsubsection{Scattering rate of HQ due to string part of the HQ potential}
The nature of the vertex for the Yukawa type interaction and the string type interaction are different from each other leaving no possibility of overlap between the two. When the string-type potential of HQ is used as the effective gluon propagator, HQ self-energy can be written as 
\begin{equation}
\Sigma(P)=i\int \frac{d^4Q}{(2\pi)^4}\frac{(\slashed{P'}+M)}{P'^2-M^2}V_S(q).
\label{HQselfs}
\end{equation}
Converting the $q_0$ integration into a summation over $q_0$, we can write
\begin{equation}
Tr\left[(\slashed{P}+M)\Sigma(P)\right]=-T\sum_{q_0}\int\frac{d^3q}{(2\pi)^3}\frac{V_S(q)}{P'^2-M^2}
Tr\left[(\slashed{P}+M)(\slashed{P'}+M)\right].
\end{equation}
where $Tr\left[(\slashed{P}+M)(\slashed{P'}+M)\right]=8M^2$.
Following the similar prescription of the spectral representation as used in the case of the Yukawa interaction, we straightway write down the scattering rate of the HQ interacting with the light partons through a string type gluon propagator in presence of a magnetic field:
\begin{align}
&\Gamma_S(E)=\frac{2\pi M^2}{E^2}\int\frac{d^3q}{(2\pi)^3}\int d\omega \rho_S(q)\nonumber\\
& \times [1+n_B(\omega)]\delta(\omega-\Vec{v}\cdot \Vec{q}),
\label{scatrateS}
\end{align}
where the string spectral function can be constructed as 
\begin{equation}
\rho_S(q)=\frac{Tm_{Dg}^2(q^2+m_s^2)}{q(q^2+m_D^2)}|V_S(q)|^2
\label{rhoS}
\end{equation}
along with the string potential written like the following
\begin{equation}
V_S(q)=\frac{16\pi\sigma}{(q^2+m_s^2)^2}\Bigg[\frac{1}{2+\frac{i\pi \sigma Tm_{Dg}^2}{q(q^2+m_D^2)(q^2+m_s^2)}}\Bigg],
\label{VSqalter}
\end{equation}
for a space-like $|\omega|<q$.

\subsection{Diffusion coefficients of HQ, $\Vec{v}\parallel \Vec{B}$}
\begin{figure}[h]
\begin{center}
\includegraphics[width=0.4\textwidth]{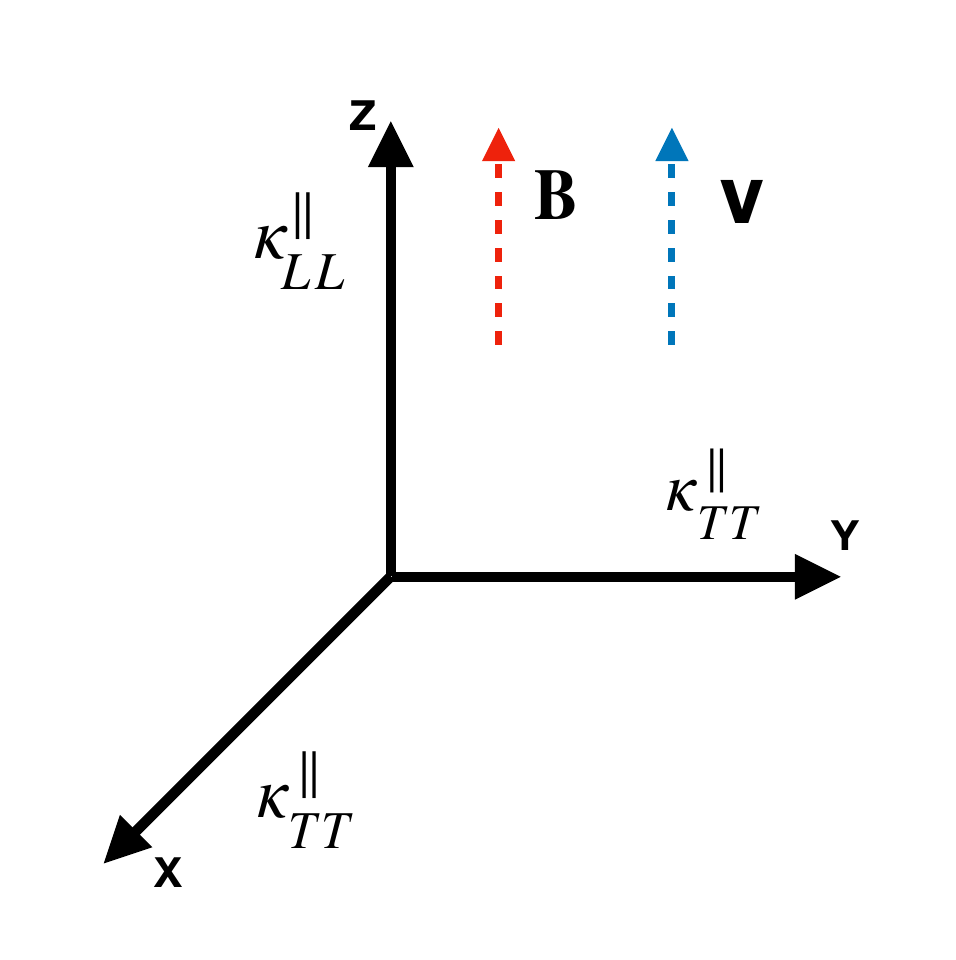}
\caption{Components of diffusion of HQ when $\Vec{v}\parallel\Vec{B}$}
\label{fig2}
\end{center}
\end{figure}

If the HQ velocity is along the direction of the magnetic field(which is taken to be the z-axis without the loss of generality), there can, possibly, be two independent directions; along the z-axis which is parallel to both $\Vec{v}$ and $\Vec{B}$ and along the x-axis or the y-axis. According to Fig.\ref{fig2}, there can be two possible independent diffusion coefficients of HQ defined by 

\begin{align}
&\kappa^{\parallel}_{LL}=\int d^3q\frac{d\Gamma}{d^3q}q_z^2,\\
&\kappa^{\parallel}_{TT}=\frac{1}{2}\int d^3q\frac{d\Gamma}{d^3q}q_T^2,
\end{align}
where $q_T^2=q_x^2+q_y^2$.
We choose a frame of reference in which $\Vec{q}$ makes an angle $\theta_q$ with the z-axis and an azimuthal angle $\phi_q$. Therefore, the delta function $\delta(\omega-\Vec{v}\cdot\Vec{q})$ can be transformed into a delta function of $\cos{\theta_q}$ which helps with the angular integration:
\begin{equation}
\delta(\omega-\Vec{v}\cdot\Vec{q})=\frac{1}{|vq|}\delta\left(\cos{\theta_q}-\frac{\omega}{vq}\right).
\label{deltavparaB}
\end{equation}
Putting the Yukawa spectral function into Eq.\ref{scatrateC} and performing the $\cos{\theta_q}$ integration with the help of the delta function given by Eq.\ref{deltavparaB}, we have the expression for the scattering rate
\begin{align}
&\Gamma_{Y}^{\parallel}(E)=\frac{g^2M^2\alpha_s Tm^2_{Dg}}{\pi E^2v}\int^{\infty}_{0}\frac{dq}{(q^2+m^2_D)^2}\nonumber\\
&\times \int d\omega [1+n_B(\omega)]
\end{align}
resulting in the expressions for the two diffusion coefficients
\begin{align}
&\kappa^{\parallel}_{LL;Y}=\frac{8\pi M^2\alpha_s^2 Tm^2_{Dg}}{ E^2v^3}\int^{\infty}_{0}\frac{dq}{(q^2+m^2_D)^2}\nonumber\\
&\times \int^{vq}_{-vq}d\omega \omega^2[1+n_B(\omega)],
\label{kparaLLc}
\end{align}
and
\begin{align}
&\kappa^{\parallel}_{TT;Y}=\frac{4\pi M^2\alpha_s^2 Tm^2_{Dg}}{ E^2v^3}\int^{\infty}_{0}\frac{dq}{(q^2+m^2_D)^2}\nonumber\\
&\times \int^{vq}_{-vq}d\omega (v^2q^2-\omega^2)[1+n_B(\omega)].
\label{kparaTTc}
\end{align}

Similarly, the string counterparts of the two diffusion coefficients come out to be
\begin{align}
&\kappa^{\parallel}_{LL;S}=\frac{\pi M^2\sigma^2 Tm_{Dg}^2}{E^2v^3}\int\frac{dq}{(q^2+m_D^2)(q^2+m_S^2)^3}\nonumber\\
&\times \int^{vq}_{-vq}d\omega \omega^2 [1+n_B(\omega)]
\label{kparaLLs}
\end{align}
and
\begin{align}
&\kappa^{\parallel}_{TT;S}=\frac{\pi M^2\sigma^2 Tm_{Dg}^2}{2E^2v^3}\int\frac{dq}{(q^2+m_D^2)(q^2+m_S^2)^3}\nonumber\\
&\times \int^{vq}_{-vq}d\omega (v^2q^2-\omega^2) [1+n_B(\omega)].
\label{kparaTTs}
\end{align}

\subsection{Diffusion coefficient of HQ, $\Vec{v}\perp\Vec{B}$}
When the velocity of HQ is perpendicular to the direction of the magnetic field, it is clear from Fig.\ref{fig3} that there can be three independent components of the diffusion tensor given by:
\begin{figure}[h]
\begin{center}
\includegraphics[width=0.4\textwidth]{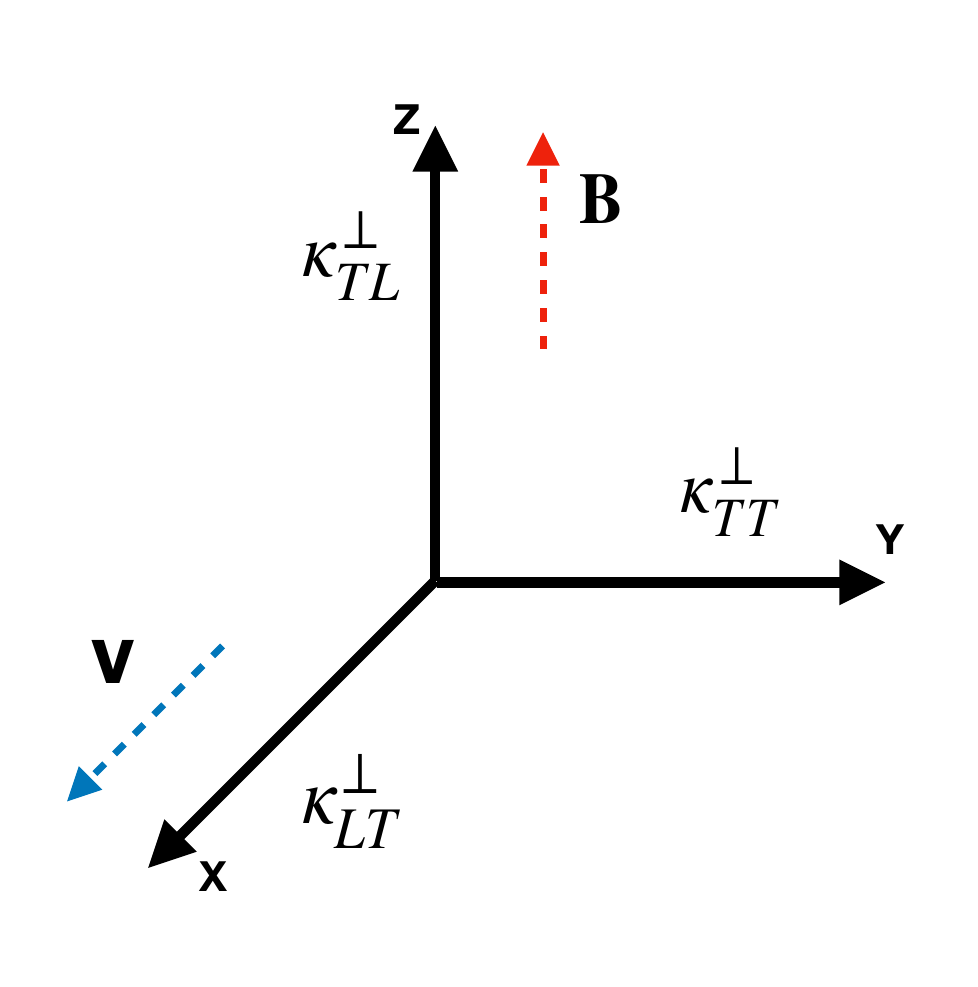}
\caption{Components of diffusion of HQ when $\Vec{v}\perp\Vec{B}$}
\label{fig3}
\end{center}
\end{figure}

\begin{align}
&\kappa^{\perp}_{LT}=\int d^3q\frac{d\Gamma}{d^3q}q_z^2,\\
&\kappa^{\perp}_{TT}=\int d^3q\frac{d\Gamma}{d^3q}q_y^2\\,
&\kappa^{\perp}_{TL}=\int d^3q\frac{d\Gamma}{d^3q}q_x^2.
\end{align}
One can assume the HQ velocity to be along the x-axis, $\Vec{v}=v\hat{x}$ with $\Vec{q}$ making an angle $\theta_q$ with the z-axis and an azimuthal angle $\phi_q$.
The delta function $\delta(\omega-\Vec{v}\cdot\Vec{q})$ can be recast in terms the azimuthal angle, $\phi_q$ to help integrate over $\phi_q$:
\begin{equation}
\delta(\omega-\Vec{v}\cdot\Vec{q})=\frac{1}{|vq\sin{\theta_q}|}\delta\left(\cos{\phi_q}-\frac{\omega}{vq\sin{\theta_q}}\right)
\label{deltavperpB}
\end{equation}
which can again be dissociated into another delta function which is a function of the angle $\phi_q$
\begin{equation}
\delta[f(\phi_q)]=\frac{\delta(\phi_q-\phi_q^0)}{|\sin{\phi_q^0}|}
\end{equation}
where $f(\phi_q)=\cos{\phi_q}-\frac{\omega}{vq\sin{\theta_q}}$ and $\phi_q^0=\cos^{-1}{\left[\frac{\omega}{vq\sin{\theta_q}}\right]}$. After some trivial algebra, we jot down the form of the scattering rate from Eq.\ref{scatrateC} as
\begin{align}
&\Gamma_Y^{\perp}(E)=\frac{g^2M^2\alpha_s Tm^2_{Dg}}{\pi E^2}\int_0^{\infty}\frac{dq}{(q^2+m_D^2)^2}\nonumber\\
&\times \int_{-1}^1d(\cos{\theta_q})\int \frac{d\omega [1+n_B(\omega)]}{\sqrt{v^2q^2\sin^2{\theta_q}}-\omega^2}
\end{align}
from which one can write down the corresponding expressions for the three diffusion coefficients
\begin{align}
&\kappa^{\perp}_{LT;Y}=\frac{4M^2\alpha_s^2 Tm^2_{Dg}}{ E^2}\int_0^{\infty}\frac{q^3dq}{(q^2+m_D^2)^2}\nonumber\\
&\times \int_{-1}^1\cos^2{\theta_q}d(\cos{\theta_q})\int\frac{d\omega [1+n_B(\omega)]}{\sqrt{v^2q^2\sin^2{\theta_q}-\omega^2}},
\label{kperpLTc}
\end{align}
\begin{align}
&\kappa^{\perp}_{TT;Y}=\frac{4M^2\alpha_s^2 Tm^2_{Dg}}{ E^2v^2}\int_0^{\infty}\frac{qdq}{(q^2+m_D^2)^2}\nonumber\\
&\times \int_{-1}^1d(\cos{\theta_q})\int d\omega [1+n_B(\omega)] \sqrt{v^2q^2\sin^2{\theta_q}-\omega^2},
\label{kperpTTc}
\end{align}
and
\begin{align}
&\kappa^{\perp}_{TL;Y}=\frac{4M^2\alpha_s^2 Tm^2_{Dg}}{ E^2v^2}\int_0^{\infty}\frac{qdq}{(q^2+m_D^2)^2}\nonumber\\
&\times \int_{-1}^1d(\cos{\theta_q})\int\frac{d\omega\omega^2[1+n_B(\omega)]}{\sqrt{v^2q^2\sin^2{\theta_q}-\omega^2}}.
\label{kperpTLc}
\end{align}

Similar procedures will result in the string counterpart of the diffusion coefficients
\begin{align}
\kappa^{\perp}_{LT;S}&=\frac{M^2\sigma^2 Tm^2_{Dg}}{E^2}\int_0^{\infty}\frac{q^3dq}{(q^2+m_D^2)(q^2+m_S^2)^3}\nonumber\\
&\times \int_{-1}^1\cos^2{\theta_q}d(\cos{\theta_q})\int\frac{d\omega[1+n_B(\omega)]}{\sqrt{v^2q^2\sin^2{\theta_q}-\omega^2}},
\label{kperpLTs}
\end{align}
\begin{align}
\kappa^{\perp}_{TT;S}&=\frac{M^2\sigma^2 Tm^2_{Dg}}{E^2v^2}\int_0^{\infty}\frac{qdq}{(q^2+m_D^2)(q^2+m_S^2)^3}\nonumber\\
&\times \int_{-1}^1d(\cos{\theta_q})\int d\omega [1+n_B(\omega)] \sqrt{v^2q^2\sin^2{\theta_q}-\omega^2},
\label{kperpTTs}
\end{align}
and
\begin{align}
\kappa^{\perp}_{TL;S}&=\frac{M^2\sigma^2 Tm^2_{Dg}}{E^2v^2}\int_0^{\infty}\frac{qdq}{(q^2+m_D^2)(q^2+m_S^2)^3}\nonumber\\
&\times \int_{-1}^1d(\cos{\theta_q})\int\frac{d\omega \omega^2[1+n_B(\omega)]}{\sqrt{v^2q^2\sin^2{\theta_q}-\omega^2}}.
\label{kperpTLs}
\end{align}

\section{Results and Discussions}\label{Results}

Calculations are performed for a charm quark of mass, $M=1.2$ GeV using a temperature dependent running coupling of QCD up to one loop
\begin{equation}
\alpha_s(T)= \frac{6\pi}{(33-2N_f)\log(2\pi T/\Lambda)}
\end{equation}
where $\Lambda=0.176~GeV$.
The string tension,$\sigma$, appearing in the string/confinement part of the HQ potential, is another temperature dependent term. This \textquote{running} string tension has been estimated using finite temperature lattice QCD.
We have taken a parametrized temperature dependent $\sigma$ from Ref.\cite{JHEP03:2022} which is calculated for a SU(2) Yang-Mills theory
\begin{equation}
\sigma(T)=\sigma_0\sqrt{1-\frac{\pi T^2}{3\sigma_0}}
\label{sigmaT}
\end{equation}
where $\sigma_0=0.450~GeV^2$ is the zero-temperature string tension. 
It is noteworthy from Eq.\ref{sigmaT} that the string tension, being attributed, principally, to the physics at lower temperature region, decreases with increasing temperature. The strength of the magnetic field is taken to be such that $eB=0.3~GeV^2\sim 15 m_{\pi}^2$.

\subsection*{Case I: $\Vec{v}\parallel \Vec{B}$}
\begin{figure}[h!]
 \centering
  \includegraphics[width=0.8\textwidth,angle=0]{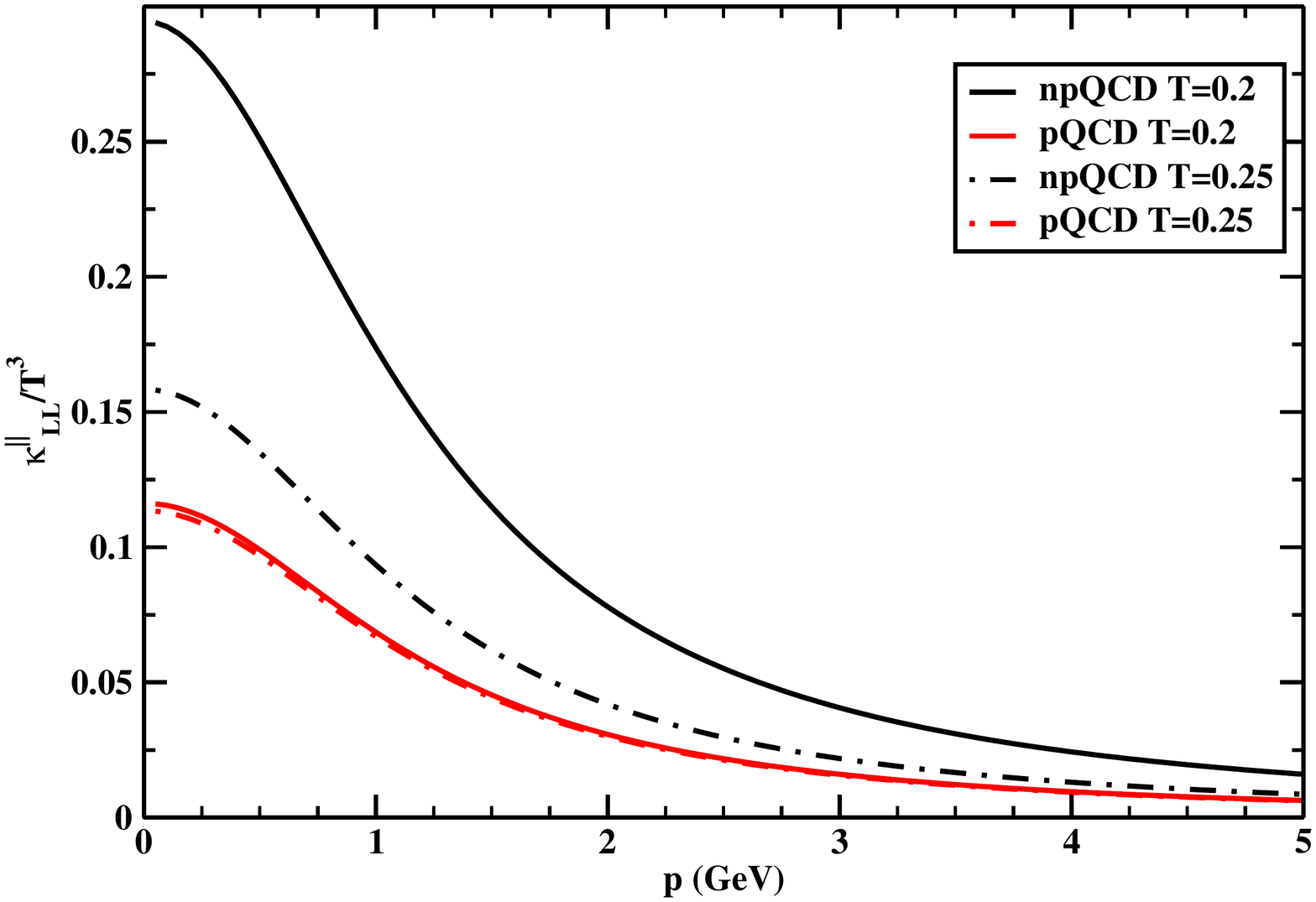}
   \caption{Comparison between npQCD and pQCD-LO results of the longitudinal diffusion for two temperatures(in GeV) with respect to momentum of charm quark}
    \label{fig:f1}
\end{figure}

Fig.\ref{fig:f1} shows the comparison between the Yukawa or pQCD and string or npQCD contributions of the charm quark diffusion coefficients, scaled by $T^3$, longitudinal to the direction of the magnetic field and the velocity of charm quark ($\Vec{v}\parallel \Vec{B}$) with respect to charm momentum at $T=200~MeV$ and $T=300~Mev$. It is evident from the plot that the non-perturbative contribution is more than the perturbative one in the low momentum region. As the momentum increases pQCD catch up with the npQCD. npQCD contributes lesser at higher temperatures and it might be possible for the pQCD part to become higher in magnitude than npQCD as is shown by the two dot-dashed lines of Fig.\ref{fig:f1}.  

\begin{figure}[h!]
 \centering
  \includegraphics[width=0.8\textwidth,angle=0]{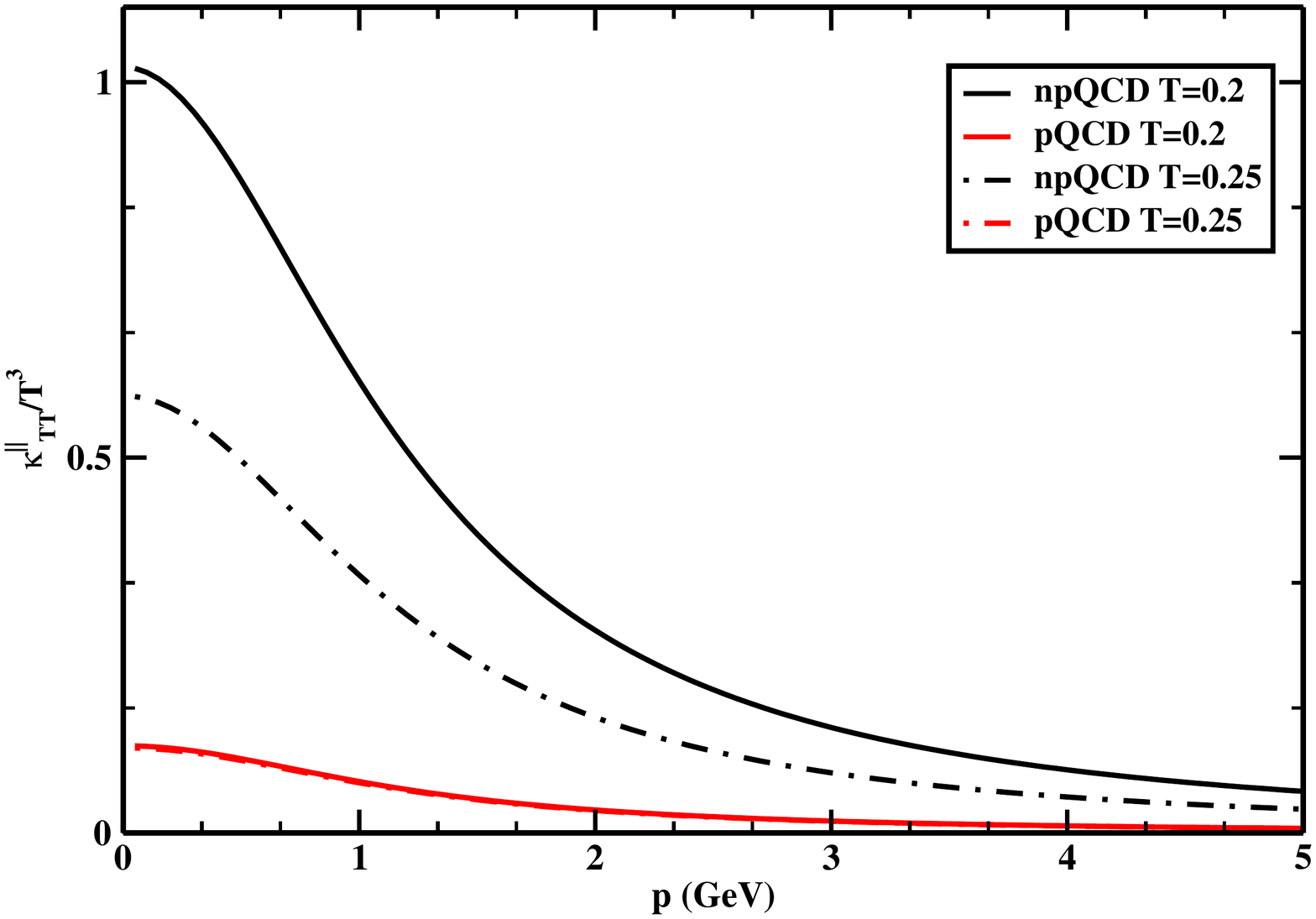}
   \caption{Comparison between npQCD and pQCD-LO results of the transverse diffusion for two temperatures(in GeV) with respect to momentum of charm quark}
    \label{fig:f2}
\end{figure}

A similar plot (Fig.\ref{fig:f2}) demonstrates the comparison between perturbative and non-perturbative contributions of charm diffusion coefficient transverse to the direction of magnetic field and the velocity of charm quark ($\Vec{v}\parallel \Vec{B}$). One can see that npQCD is larger than the pQCD contribution in lower momenta of charm and as the momentum is increasing, pQCD is becoming similar in magnitude to the npQCD counterpart.
\begin{figure}[h!]
 \centering
  \includegraphics[width=0.8\textwidth,angle=0]{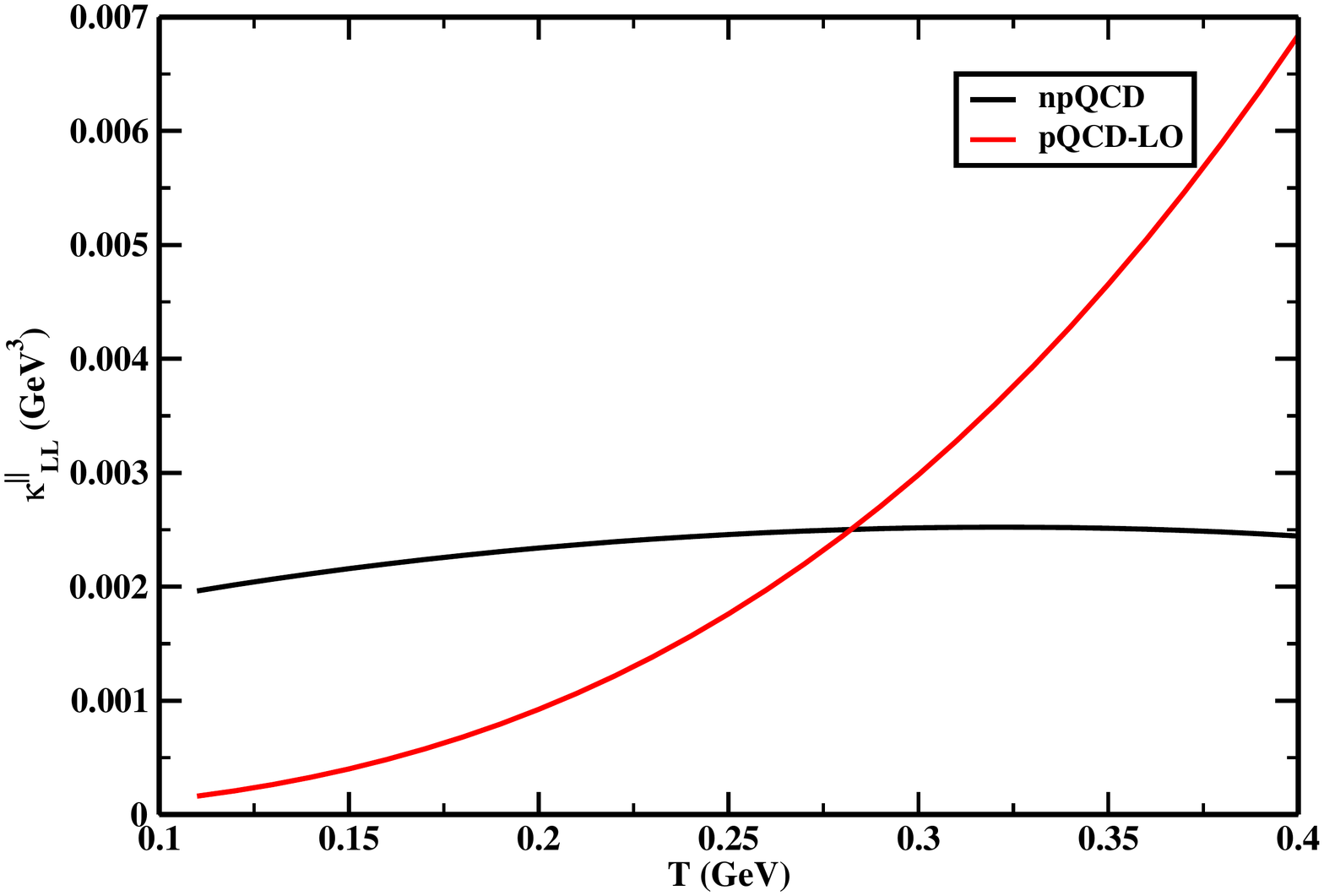}
   \caption{Comparison between the temperature variation of npQCD and pQCD-LO results of the longitudinal diffusion for a charm quark of 0.1 $GeV$ momentum.}
    \label{fig:f2a}
\end{figure}
One can observe from Fig.~\ref{fig:f2a} that npQCD contributes more than pQCD at lower temperatures until a particular temperature is reached beyond which pQCD starts to dominate over pQCD. This is just a generic plot describing the component of diffusion that is longitudinal both to the charm velocity and magnetic field when the momentum of the charm quark in 1 $GeV$. A similar plot follows for the transverse component as well.  

\subsection*{Case II: $\Vec{v}\perp \Vec{B}$}

\begin{figure}[h!]
 \centering
  \includegraphics[width=0.8\textwidth,angle=0]{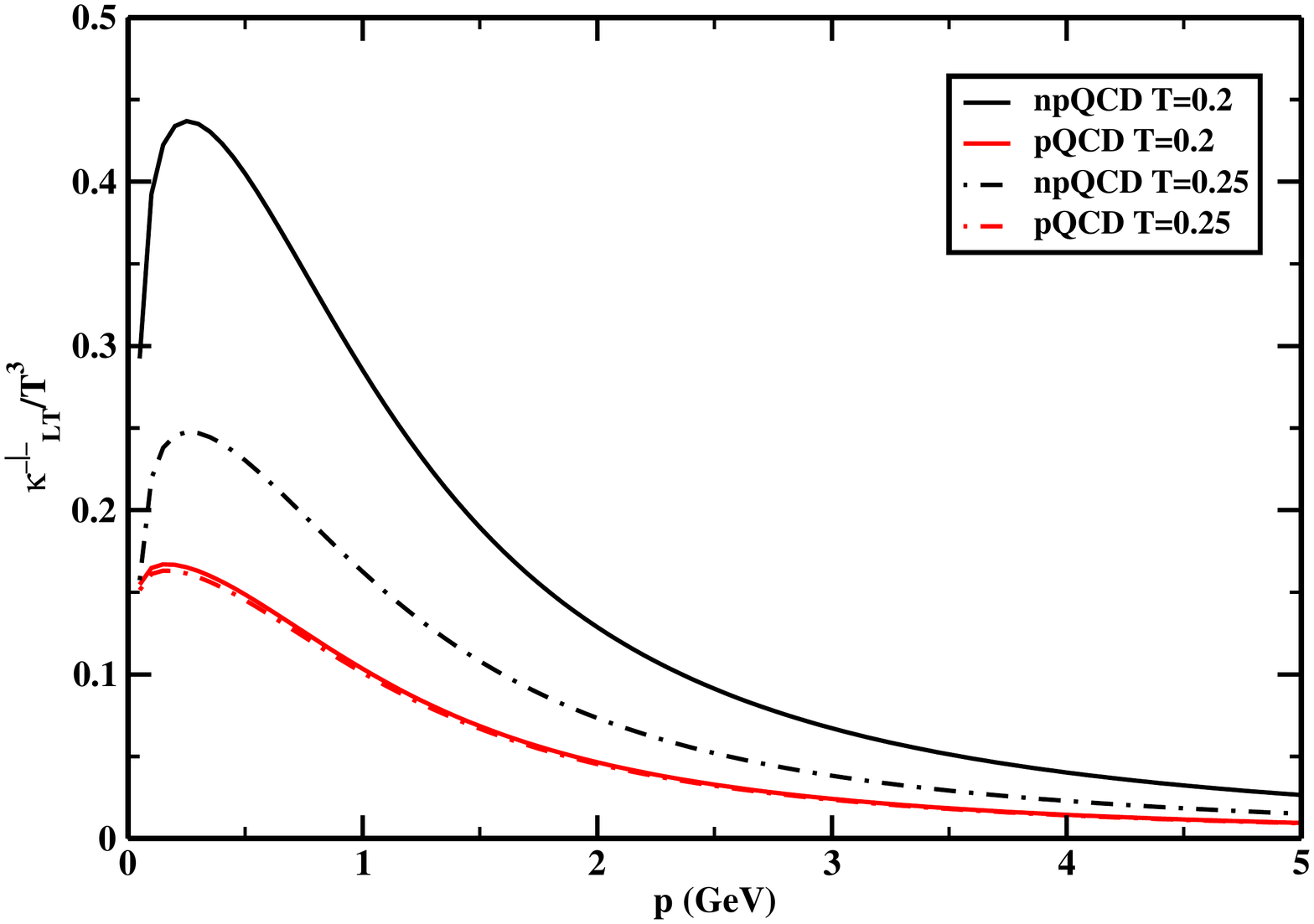}
   \caption{Comparison between npQCD and pQCD-LO results of the diffusion transverse to the charm velocity and longitudinal to the magnetic field for two temperatures(in GeV) with respect to momentum of charm quark}
    \label{fig:f3}
\end{figure}

\begin{figure}[h!]
 \centering
  \includegraphics[width=0.8\textwidth,angle=0]{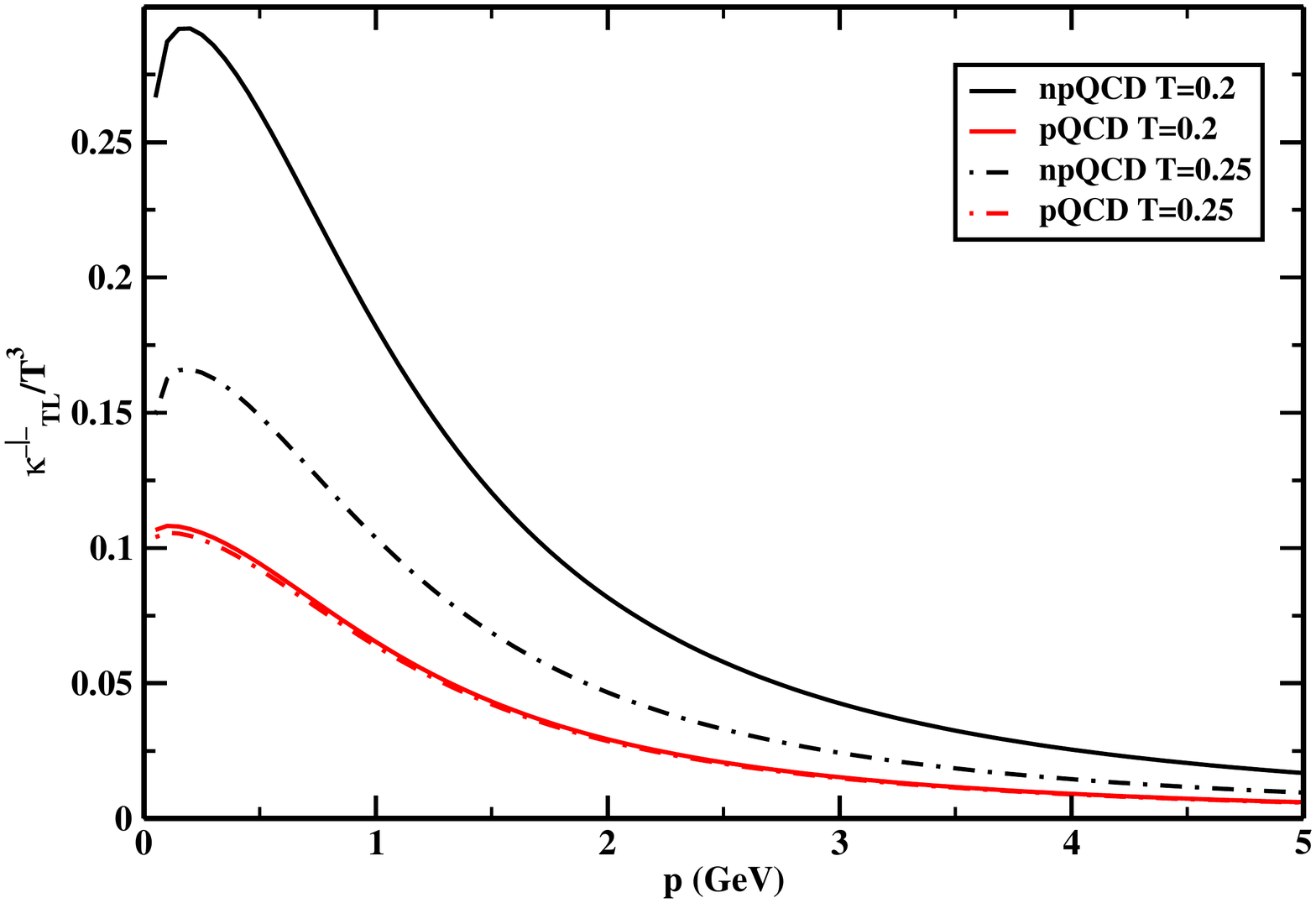}
   \caption{Comparison between npQCD and pQCD-LO results of the diffusion longitudinal to the charm velocity and transverse to the magnetic field for two temperatures(in GeV) with respect to momentum of charm quark}
    \label{fig:f4}
\end{figure}
\begin{figure}[h!]
 \centering
  \includegraphics[width=0.8\textwidth,angle=0]{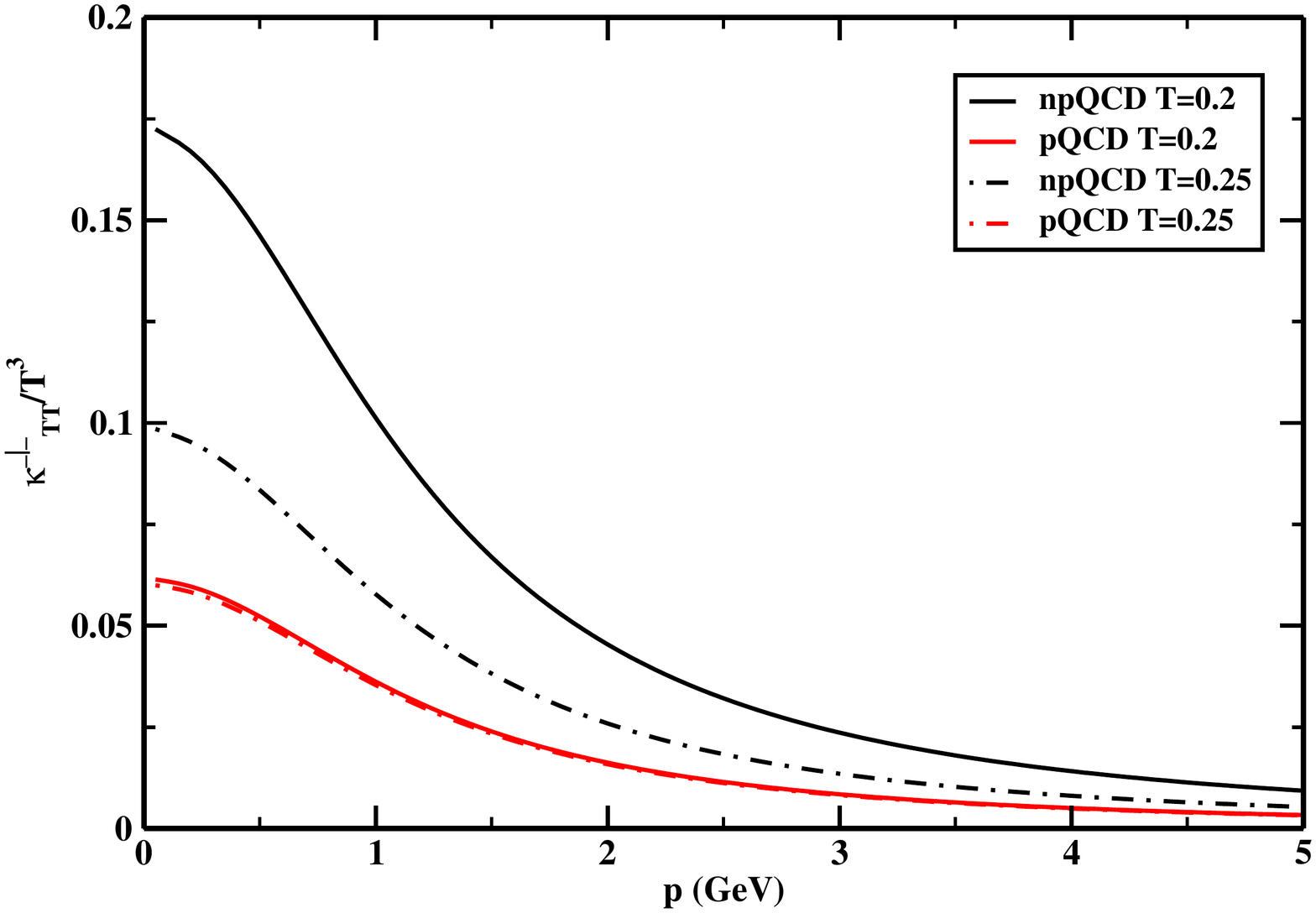}
   \caption{Comparison between npQCD and pQCD-LO results of the diffusion transverse to both the charm velocity and the magnetic field for two temperatures(in GeV) with respect to momentum of charm quark}
    \label{fig:f5}
\end{figure}

Figs.\ref{fig:f3}, \ref{fig:f4} and \ref{fig:f5} show the comparison between the npQCD and LO pQCD calculations of the three charm quark diffusion components when $\Vec{v}\perp \Vec{B}$ for two temperatures(in GeV) plotted with respect to the momentum of the charm quark. As expected, it is observed that at $T=200~MeV$ temperature, npQCD exceeds pQCD contribution at lower momentum while npQCD effects reduce as momentum increases. At a higher temperature, however, pQCD is higher than npQCD but in a very very small amount. This is because of the decreasing effect of npQCD at higher temperatures manifested by the form of string tension, $\sigma(T)$ taken in the present work specifically.  

\begin{figure}[h!]
 \centering
  \includegraphics[width=0.8\textwidth,angle=0]{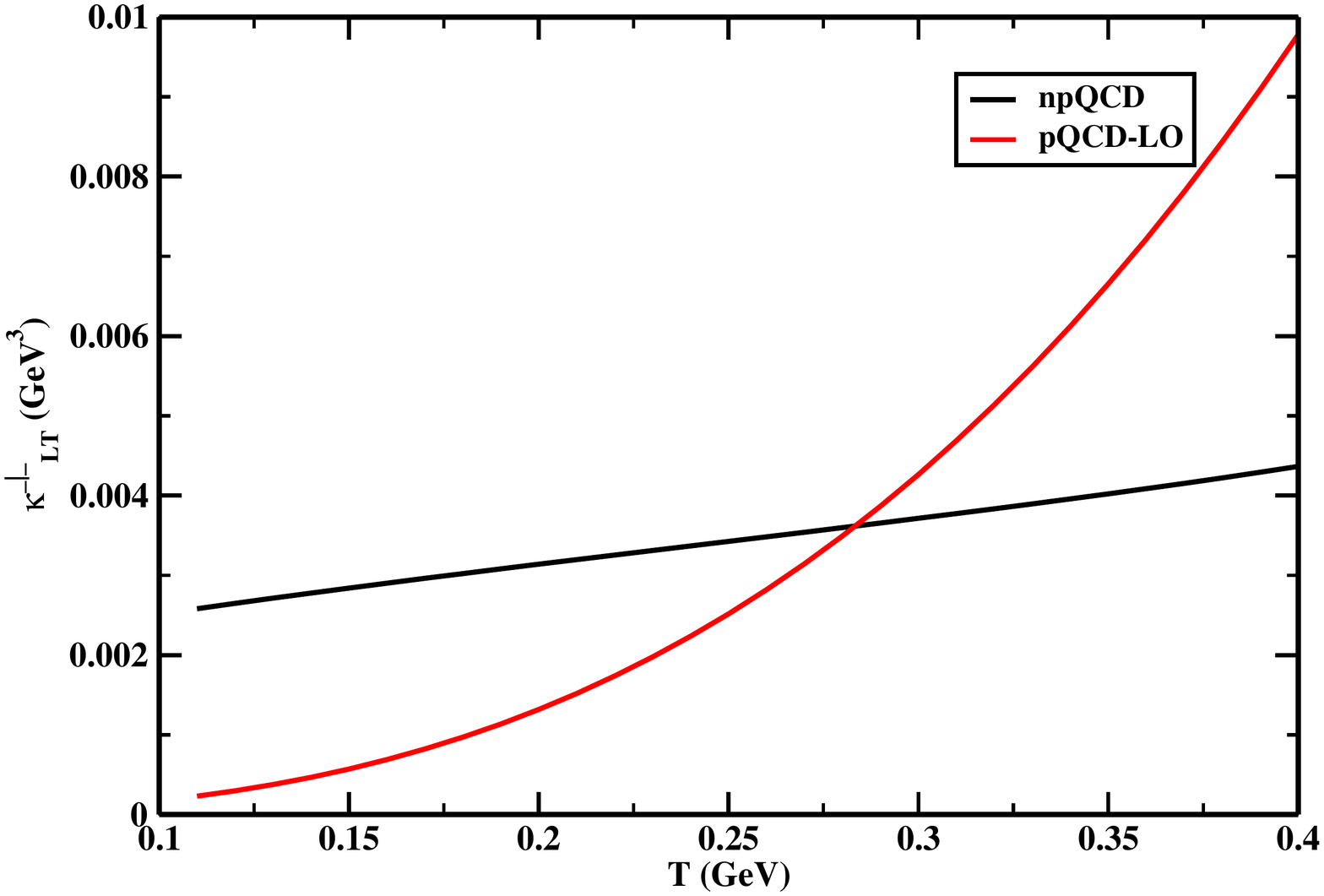}
   \caption{Temperature variation of $\kappa_{LT}^{\perp}$ for npQCD and pQCD for a charm quark with momentum 0.1 $GeV$.}
    \label{fig:f5a}
\end{figure}

It is useful to show the temperature dependence of at least one component of the diffusion for $\Vec{v}\perp \Vec{B}$ case to understand the relative magnitudes of the pQCD and the npQCD contributions. Fig.~\ref{fig:f5a} depicts the variation of the component of the diffusion that is longitudinal to the magnetic field while being transverse to the charm velocity with the temperature. The nature of the plots for the other two components of the diffusion is similar. npQCD is more significant than pQCD at lower temperatures and becomes less and less important as the temperature crosses a particular value. The crossover temperature will depend upon the competition between the decreasing running coupling and decreasing string tension with increasing temperature.

\begin{figure}[h!]
 \centering
  \includegraphics[width=0.8\textwidth,angle=0]{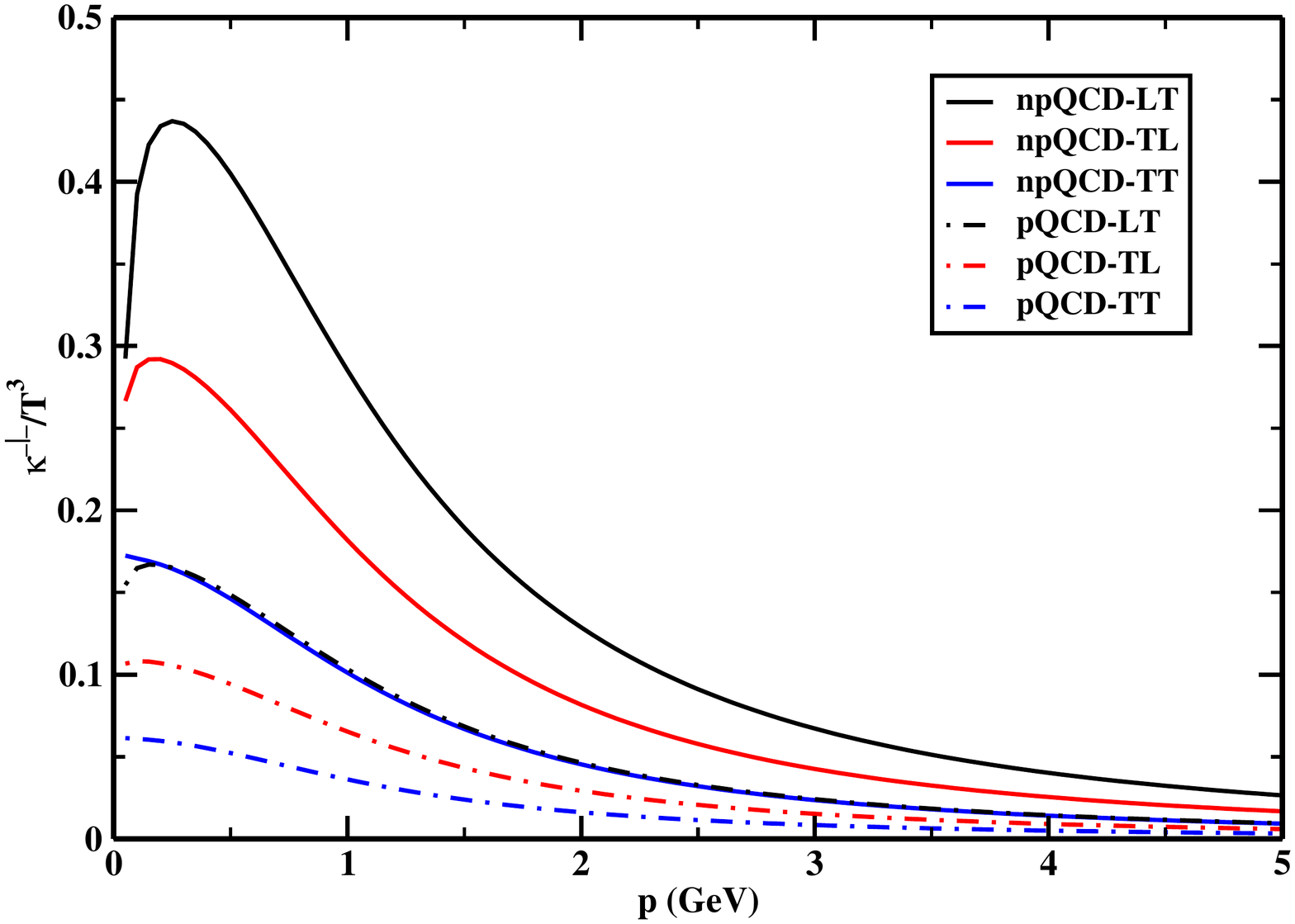}
   \caption{Hierarchy among the three components of charm diffusion for $\Vec{v}\perp \Vec{B}$ with respect to charm momentum at $T=0.2 GeV$}
    \label{fig:f6}
\end{figure}

Fig.\ref{fig:f6} illustrates the hierarchy among the three diffusion coefficients of charm quark as a function of momentum when $\Vec{v}\perp \Vec{B}$ for both perturbative and non-perturbative interactions. The two kinds of anisotropy in the system due to the presence of the magnetic field(along the z-direction) and the velocity of charm quark(along the x-direction) are reflected in the diffusion coefficients obtained in three different directions, i.e. $\kappa^{\perp}_{LT}>\kappa^{\perp}_{TL}>\kappa^{\perp}_{TT}$. Therefore, a charm quark jet moving in the direction perpendicular to that of the magnetic field would experience more broadening in the direction longitudinal to the magnetic field than in the reaction plane. It is intriguing to compare the result with Ref.~\cite{PRC75:2007} in which a similar hierarchy has been observed much as the source of anisotropy considered in that treatise was of a different nature.

\section{Summary and Outlook}\label{Summary}

The present work puts forward a methodology to estimate the diffusion coefficients of heavy quarks in the non-pQCD domain within the scope of the pQCD framework. The diffusion coefficients are calculated for a charm quark with mass 1.2 $GeV$ moving with velocity i) parallel and ii) perpendicular to the direction of the magnetic field, respectively. Both cases find the string or npQCD contribution to be higher than the pQCD contribution at lower temperatures and lower momentum range of the charm quark. As temperature and charm quark momentum increases the non-perturbative effects start reducing and perturbative contribution begins to gain on.

We intend explore, in the next phase of our work, the connection of the anisotropic nature of the HQ diffusion coefficients due to the presence of the magnetic field with the pertinent experimental observations. It is established that pQCD is not successful in explaining the phenomenology of HQ at low to intermediate $p_T$ region as well as at lower temperatures of the medium, and thus it necessitates the study of the HQ transport in the npQCD regime.  
One would see from the magnitudes of the diffusion coefficients that when they are used as inputs to solve the HQ transport equation like Langevin or Fokker-Planck, npQCD contributes more than pQCD to energy loss or nuclear modification,$R_{AA}$, elliptic flow, $v_2$ or jet quenching parameter, $\hat{q}$ of charm quark at lower temperatures and momentum. We intend to focus on these crucial aspects concerning heavy quarks in hot QCD medium.

It is important to emphasize that the two distinctive properties of QCD -$\textit{viz}$, i) confinement - designated by the magnitude of the string tension and ii) asymptotic freedom featured by the running coupling, play a pivotal role in determining the evolution of HQ transport with temperature. As a matter of course, the results presented in this work are based on an effective way to describe the situation. In spite of various limitations and approximations, our approach provides a new theoretical framework that can be propelled forward, in the near future, for more realistic results.  

Diverse extensions and improvements of the present work can be realized as future endeavors. We plan to broaden our investigation to encompass scenarios involving a medium to a weak magnetic field, a more realistic condition. In this context, one has to allow for the higher Landau Levels to play a significant role in all theoretical estimations. Most of all, as mentioned before, we will attempt to explain the experimental realization of HQ transport in terms of the observables, such as $R_{AA}$, $v_2$, and so forth in the new light of the non-perturbative physics.

\acknowledgments
V.C. and S.K.D. acknowledge the SERB, Dept. of Science and Technology, Govt. of India for the  Core Research Grant (CRG) [CRG/2020/002320]. S. M. acknowledges the Indian Institute of Technology Gandhinagar for the institute postdoctoral fellowship.

\appendix

\section{Calculation of the Yukawa scattering rate:}

\begin{align}
Tr\left[(\slashed{P}+M)\Sigma(P)\right]&=-g^2T\sum_{q_0}\int \frac{d^3q}{(2\pi)^3}\frac{V_Y(q)}{P'^2-M^2}\nonumber\\
&\times Tr\left[(\slashed{P}+M)\gamma_{\mu}(\slashed{P'}+M)\gamma^{\mu}\right].
\label{traceC1}
\end{align}
The trace inside the integration in Eq.\ref{traceC1} is found to be
\begin{align}
Tr\left[(\slashed{P}+M)\gamma_{\mu}(\slashed{P'}+M)\gamma^{\mu}\right]&=\left(M^2+P.Q\right)\nonumber\\
&=8M^2,
\end{align}
as $P.Q=E\omega-E\Vec{v}\cdot\Vec{q}=0$ with $\omega=\Vec{v}\cdot\Vec{q}$ for small momentum transfer. Therefore, Eq.\ref{traceC1} becomes
\begin{equation}
Tr\left[(\slashed{P}+M)\Sigma(P)\right]=-8g^2M^2T\sum_{q_0}\int\frac{d^3q}{(2\pi)^3}\frac{V_Y(q)}{P'^2-M^2}.
\label{traceC2}
\end{equation}
The standard way to solve Eq.\ref{traceC2} is to introduce a spectral representation of the HQ and gluon propagators as the following:
\begin{align}
\frac{1}{P'^2-M^2}&=-\frac{1}{2E'}\int^{1/T}_{0}d\tau'e^{(p_0-q_0)\tau'}\nonumber\\
&\times\left[(1-n_F(E'))e^{-E'\tau'}-n_F(E')e^{E'\tau'}\right]
\label{HQsaclay}
\end{align}
and
\begin{equation}
V_Y(q)=-\int^{1/T}_{0}d\tau e^{q_0\tau}\int^{\infty}_{-\infty}d\omega \rho_Y(q)[1+n_B(\omega)]e^{-\omega \tau}.
\label{gluonsaclay}
\end{equation}
Putting Eq.\ref{HQsaclay} and Eq.\ref{gluonsaclay} into Eq.\ref{traceC2} and using the identity
\begin{equation}
T\sum_{q_0}e^{q_0(\tau-\tau')}=\delta(\tau-\tau')
\label{delta}
\end{equation}
to perform $\tau'$-integration, one gets
\begin{align}
&Tr\left[(\slashed{P}+M)\Sigma(P)\right]=-\frac{4g^2M^2}{E}\int\frac{d^3q}{(2\pi)^3}\int^{\infty}_{-\infty}d\omega \rho_Y(q)\nonumber\\
&\times [1+n_B(\omega)]\Bigg[(1-n_F(E'))\int d\tau e^{(p_0-E'-\omega)\tau}\nonumber\\
&-n_F(E')\int d\tau e^{(p_0+E'-\omega)\tau}\Bigg].
\label{traceC3}
\end{align}
$\tau$-integration gives 
\begin{equation}
\int^{1/T}_{0}d\tau e^{(p_0\mp E'-\omega)\tau}=\frac{e^{(p_0\mp E'-\omega)/T}-1}{p_0\mp E'-\omega}.
\end{equation}
To calculate the scattering rate we are to take the imaginary part of the HQ self-energy amounting to the following identities:
\begin{equation}
\Im{\left(\frac{1}{p_0\mp E'-\omega+i\epsilon}\right)}=-\pi \delta(p_0\mp E'-\omega).
\end{equation}
As the term involving $\delta(p_0+E'-\omega)$ vanishes as $\omega\sim T$, the only term worth taking is that consisting of $\delta(p_0-E'-\omega)$. Dropping the exponentially suppressed Fermi-Dirac distribution function of the HQ and noting the fact that for $p_0=(2n+1)i\pi T$, $e^{p_0/T}=1$, we get
\begin{align}
&Tr\left[(\slashed{P}+M)\Im{\Sigma(P)}\right]=-\frac{4\pi g^2 M^2}{E}(1+e^{-E/T})\int\frac{d^3q}{(2\pi)^3}\nonumber\\
&\int^{\infty}_{-\infty}d\omega \rho_Y(q)[1+n_B(\omega)]\delta(\omega-\Vec{v}\cdot \Vec{q}),
\label{traceC4}
\end{align}
in which $\rho_Y(q)=-\frac{\Im{V_Y(q)}}{\pi}$. When $|\omega|\sim T$, it is possible to write down the Yukawa spectral function as 
\begin{equation}
\rho_Y(q)=\frac{Tm_{Dg}^2}{q(4\pi \alpha_s)}|V_Y(q)|^2,
\label{rhoY}
\end{equation}
if the Yukawa potential is represented in the following form for $|\omega|<q$
\begin{equation}
V_Y(q)=\frac{4\pi \alpha_s}{(q^2+m_D^2)+i\pi\frac{T}{q}m_{Dg}^2}.
\label{VYqalter}
\end{equation}




\begin{thebibliography}{99}

\bibitem{Prino:2016cni} 
  F.~Prino and R.~Rapp,
  J.\ Phys.\ G {\bf 43}, no. 9, 093002 (2016)
  
\bibitem{Andronic:2015wma} 
  A.~Andronic {\it et al.},
  Eur.\ Phys.\ J.\ C {\bf 76}, no. 3, 107 (2016)

\bibitem{Rapp:2018qla}
R.~Rapp, P.~B.~Gossiaux, A.~Andronic, R.~Averbeck, S.~Masciocchi, A.~Beraudo, E.~Bratkovskaya, P.~Braun-Munzinger, S.~Cao and A.~Dainese, \textit{et al.}
Nucl. Phys. A \textbf{979}, 21-86 (2018)
\bibitem{Aarts:2016hap} 
  G.~Aarts {\it et al.},
  Eur.\ Phys.\ J.\ A {\bf 53}, no. 5, 93 (2017)
\bibitem{Cao:2018ews}
S.~Cao, G.~Coci, S.~K.~Das, W.~Ke, S.~Y.~F.~Liu, S.~Plumari, T.~Song, Y.~Xu, J.~Aichelin and S.~Bass, \textit{et al.}
Phys. Rev. C \textbf{99}, no.5, 054907 (2019)
\bibitem{Dong:2019unq}
X.~Dong and V.~Greco,
Prog. Part. Nucl. Phys. \textbf{104}, 97-141 (2019)
\bibitem{Das:2022lqh}
S.~K.~Das, P.~Palni, J.~Sannigrahi, J.~e.~Alam, C.~W.~Aung, Y.~Bailung, D.~Banerjee, G.~G.~Barnaf\"oldi, S.~C.~Behera and P.~P.~Bhaduri, \textit{et al.}
[arXiv:2208.13440 [nucl-th]].


\bibitem{IJMP24:2009}
V. Skokov, A. Y. Illarionov, and V. Toneev, Int. J. Mod. Phys. A {\bf 24}, 5925 (2009).

\bibitem{PRC83:2011_1}
V. Voronyuk, V. D. Toneev, W. Cassing, E. L. Bratkovskaya,
V. P. Konchakovski, and S. A. Voloshin, Phys. Rev. C {\bf 83},
054911 (2011).

\bibitem{PLB710:2012}
A. Bzdak and V. Skokov, Phys. Lett. B {\bf 710}, 171 (2012).

\bibitem{PRC85:2012}
W. T. Deng and X. G. Huang, Phys. Rev. C {\bf 85}, 044907 (2012).

\bibitem{PLB718:2013}
J. Bloczynski, X. G. Huang, X. Zhang and J. Liao, Phys. Lett. B {\bf 718}, 1529-1535 (2013).

\bibitem{PRL110:2013}
A. Bzdak and V. Skokov, Phys. Rev. Lett. {\bf 110}, 192301 (2013).

\bibitem{PRC88:2013}
K. Tuchin, Phys. Rev. C {\bf 88}, 024911 (2013).

\bibitem{NPA929:2014}
L. McLerran and V. Skokov, Nucl. Phys. A {\bf 929}, 184, (2014).

\bibitem{PRC93:2016}
K. Tuchin, Phys. Rev. C {\bf 93}, 014905 (2016).

\bibitem{NPA803:2008}
D. E. Kharzeev, L. D. McLerran, and H. J. Warringa, Nucl.
Phys. A {\bf 803}, 227 (2008).

\bibitem{PRD78:2008}
K. Fukushima, D. E. Kharzeev,
and H. J. Warringa, Phys. Rev. D {\bf 78}, 074033 (2008).

\bibitem{PRL107:2011}
Y. Burnier, D. E. Kharzeev, J. Liao, and H. U. Yee, Phys.
Rev. Lett. {\bf 107}, 052303 (2011).

\bibitem{PRD83:2011}
E. V. Gorbar, V. A. Miransky, and I. A. Shovkovy, Phys.
Rev. D {\bf 83}, 085003 (2011).

\bibitem{PRC83:2011_2}
K. Tuchin, Phys. Rev. C {\bf 83}, 017901 (2011), 

\bibitem{PLB710-230:2012}
A. A. Andrianov, V. A. Andrianov, D. Espriu, and X.
Planells, Phys. Lett. B {\bf 710}, 230 (2012).

\bibitem{PRL109:2012}
G. Basar, D. Kharzeev, and V. Skokov, Phys. Rev. Lett. {\bf 109},
202303 (2012).

\bibitem{PRD86:2012}
K. Fukushima and K. Mameda, Phys. Rev. D {\bf 86}, 071501
(2012).

\bibitem{PRD88:2013_1}
H. U. Yee, Phys. Rev. D {\bf 88}, 026001 (2013).

\bibitem{PRD89:2014_1}
B. Muller, S. Y. Wu, and D. L. Yang, Phys. Rev. D {\bf 89},
026013 (2014).

\bibitem{PRC90:2014_1}
Y. Yin, Phys. Rev. C {\bf 90}, 044903 (2014).


\bibitem{Das:2016cwd}
S.~K.~Das, S.~Plumari, S.~Chatterjee, J.~Alam, F.~Scardina and V.~Greco,
Phys. Lett. B \textbf{768} (2017), 260-264

\bibitem{STAR:2019clv}
J.~Adam \textit{et al.} [STAR],
Phys. Rev. Lett. \textbf{123} (2019) no.16, 162301

\bibitem{ALICE:2019sgg}
S.~Acharya \textit{et al.} [ALICE],
Phys. Rev. Lett. \textbf{125} (2020) no.2, 022301


\bibitem{PRD88:2013_2}
C. S. Machado, F. S. Navarra, E. G. de Oliveira, J. Noronha,
and M. Strickland, Phys. Rev. D {\bf 88}, 034009 (2013).

\bibitem{PRD89:2014_2}
C. S. Machado, S. I. Finazzo, R. D. Matheus, and J.
Noronha, Phys. Rev. D {\bf 89}, 074027 (2014).

\bibitem{PRD93:2016_1}
P. Gubler, K. Hattori, S. H. Lee, M. Oka, S. Ozaki, and K.
Suzuki, Phys. Rev. D {\bf 93}, 054026 (2016).

\bibitem{PRD93:2016_2}
K. Fukushima, K. Hattori, H. Yee and Y. Yin, Phys. Rev. D {\bf 93}, 074028 (2016).

\bibitem{PRD100:2019}
M. Kurian, S. K. Das and V. Chandra, Phys. Rev. D {\bf 100}, 074003 (2019).

\bibitem{Kurian:2020kct}
M.~Kurian, V.~Chandra and S.~K.~Das,
Phys. Rev. D \textbf{101} (2020) no.9, 094024

\bibitem{JHEP05:2020}
B. Singh, S. Mazumder and H. Mishra, JHEP {\bf 05}, 068 (2020).

\bibitem{arxiv2004:2020}
B. Singh, M. Kurian, S. Mazumder, H. Mishra, V. Chandra and S. K. Das, arXiv:2004.11092[hep-ph] (2020).

\bibitem{PRD105:2022}
A. Bandyopadhyay, J. Liao, and H. Xing
Phys. Rev. D {\bf 105}, 114049 (2022). 

\bibitem{JHEP02:2008}
S. Caron-Huot and G. D. Moore, JHEP {\bf 02}, 081 (2008).






\bibitem{vanHees:2005wb}
H.~van Hees, V.~Greco and R.~Rapp,
Phys. Rev. C \textbf{73} (2006), 034913

\bibitem{PRL100:2008}
H. van Hees, M. Mannarelli, V. Greco, and R. Rapp,
Phys. Rev. Lett. {\bf 100}, 192301 (2008).


\bibitem{Gossiaux:2008jv}
P.~B.~Gossiaux and J.~Aichelin,
Phys. Rev. C \textbf{78} (2008), 014904

\bibitem{Gossiaux:2009mk}
P.~B.~Gossiaux, R.~Bierkandt and J.~Aichelin,
Phys. Rev. C \textbf{79} (2009), 044906

\bibitem{Das:2012ck}
S.~K.~Das, V.~Chandra and J.~e.~Alam,
J. Phys. G \textbf{41} (2013), 015102

\bibitem{PRC86:2012}
M. He, R. J. Fries, and R. Rapp, Phys. Rev. C {\bf 86}, 014903
(2012).

\bibitem{Berrehrah:2013mua}
H.~Berrehrah, E.~Bratkovskaya, W.~Cassing, P.~B.~Gossiaux, J.~Aichelin and M.~Bleicher,
Phys. Rev. C \textbf{89} (2014) no.5, 054901

\bibitem{PLB747:2015}
S. K. Das, F. Scardina, S. Plumari, and V. Greco, Phys.
Lett. B {\bf 747}, 260 (2015).

\bibitem{Scardina:2017ipo}
F.~Scardina, S.~K.~Das, V.~Minissale, S.~Plumari and V.~Greco,
Phys. Rev. C \textbf{96} (2017) no.4, 044905

\bibitem{Song:2015sfa}
T.~Song, H.~Berrehrah, D.~Cabrera, J.~M.~Torres-Rincon, L.~Tolos, W.~Cassing and E.~Bratkovskaya,
Phys. Rev. C \textbf{92} (2015) no.1, 014910

\bibitem{Beraudo:2014boa}
A.~Beraudo, A.~De Pace, M.~Monteno, M.~Nardi and F.~Prino,
Eur. Phys. J. C \textbf{75} (2015) no.3, 121

\bibitem{EPJC78:2018_1}
S. Plumari, V. Minissale, S. K. Das, G. Coci,
and V. Greco, Eur. Phys. J. C {\bf 78}, 348 (2018).

\bibitem{EPJC78:2018_2}
J. Song, H.-H. Li, and F.-L. Shao, Eur. Phys. J. C {\bf 78},
344 (2018).

\bibitem{PRC101:2020}
S. Cho, K.-J. Sun, C. M. Ko, S. H. Lee, and Y. Oh, Phys.
Rev. C {\bf 101}, 024909 (2020).

\bibitem{PRC94:2016}
S. Cao, T. Luo, G.-Y. Qin, and X.-N. Wang, Phys. Rev.
C {\bf 94}, 014909 (2016).

\bibitem{Xu:2017obm}
Y.~Xu, J.~E.~Bernhard, S.~A.~Bass, M.~Nahrgang and S.~Cao,
Phys. Rev. C \textbf{97} (2018) no.1, 014907

\bibitem{Liu:2018syc}
S.~Y.~F.~Liu, M.~He and R.~Rapp,
Phys. Rev. C \textbf{99} (2019) no.5, 055201

\bibitem{gribov:najmul}
S. Madni, A. Mukherjee, A. Bandyopadhyay and N. Haque,
Phys. Lett. B \textbf{838}, 137714.

\bibitem{arxiv2112:2021}
W. Xing, G. Qin and S. Cao, arXiv:2112.15062[hep-ph] (2021).

\bibitem{PRD97:2018}
B. Singh, L. Thakur and H. Mishra, Phys. Rev. D {\bf 97}, 096011 (2018).

\bibitem{JHEP03:2022}
F. Caristo, M. Caselle, N. Magnoli, A. Nada, M. Panero and A. Smecca, JHEP {\bf 03}, 115 (2022).

\bibitem{PRC75:2007}
P. Romatschke, Phys. Rev. C {\bf 75}, 014901 (2007).

\end{thebibliography}
\end{document}